\documentclass[prb,aps,superscriptaddress,twocolumn]{revtex4}

\usepackage{graphicx}
\usepackage{dcolumn}
\usepackage{bm}
\usepackage{amssymb}

\begin{document}


\title{Large-$J$ approach to strongly coupled spin-orbital systems}

\author{Gia-Wei Chern}
\affiliation{Department of Physics, University of Wisconsin,
Madison, Wisconsin 53706, USA}
\author{Natalia Perkins}
\affiliation{Department of Physics, University of Wisconsin,
Madison, Wisconsin 53706, USA}

\date{\today}

\begin{abstract}
We present a novel approach to study the ground state and elementary
excitations in compounds where spins and orbitals are entangled by
on-site relativistic spin-orbit interaction. The appropriate degrees
of freedom are localized states with an effective angular momentum
$J$. We generalize $J$ to arbitrary large values while maintaining
the delicate spin-orbital entanglement. After projecting the
inter-site exchange interaction to the manifold of effective spins,
a systematic $1/J$ expansion of the effective Hamiltonian is
realized using the Holstein-Primakoff transformation. Applications
to representative compounds Sr$_2$IrO$_4$ and particularly vanadium
spinels $A$V$_2$O$_4$ are discussed.
\end{abstract}

\maketitle

Transition metal compounds with partially filled orbitals have
attracted considerable attention in recent years. \cite{tokura00}
The interplay of spin and orbital degrees of freedom leads to a
variety of interesting ground states and elementary excitations.
Particularly, a long-range orbital order in the ground state makes
possible the propagating orbital excitations, or orbitons,
\cite{forte08} in addition to the familiar spin waves. The intricate
spin-orbital interaction dictated  by the Kugel-Khomskii \cite{kk}
type superexchange (SE) Hamiltonian gives rise to a magnon spectrum
which depends on the orbital order, and vice versa. Meanwhile,
interaction between the two types of quasiparticles leads to their
scattering and decay.

Recently, there is a growing number of orbitally degenerate
compounds whose elementary excitations can not be described by the
above paradigm based on magnons and orbitons. A common feature
shared by these compounds is the presence of a strong relativistic
spin-orbit (SO) interaction $V_{LS} = \lambda(\mathbf L\cdot\mathbf
S)$; most notable among them are vanadium and iridium oxides
containing V$^{3+}$ and Ir$^{4+}$ ions, respectively.
\cite{lee04,reehuis03,garlea08,kim08}  To understand these systems,
one should start with localized degrees of freedom which diagonalize
the SO coupling. \cite{ot04,kim08,jackeli08} Such atomic states are
usually composed of complex orbitals and are characterized by an
effective angular momentum $J$. The on-site spin-orbital
entanglement leads not only to novel ground states with
orbital-moment ordering, \cite{ot04,horsch03} but also to the
appearance of {\em new} types of quasiparticles which carry both
spin and orbital characters.

Although several theoretical studies have been devoted to
understanding the ground-state properties in the limit of strong SO
coupling, \cite{jackeli08,ot04,dimatteo05,maitra07,horsch03} very
few works address the problem of elementary excitations.
A straightforward approach is to use the magnon/orbiton description
starting from the $\lambda=0$ ground state and to treat SO
interaction as a perturbation. \cite{oles07} However, as the
large-$\lambda$ ground state is fundamentally different from that of
$\lambda=0$, such a formulation would be difficult to capture
features like ordering of orbital moments. Another alternative is
the magnetic-exciton model.
Its application to vanadium spinels reveals an excitation spectrum
characterized by transitions between states with different
$J_z$.\cite{perkins07} Despite its success in providing a basic
structure of excitations in the large $\lambda$ limit, the
magnetic-exciton model relies on an ill-controlled mean-field
decoupling of the SE interaction. Besides, the model itself does not
offer much insight to the ground-state structure.

In this paper, we propose a new theoretical framework to study
magnetically ordered ground state and its quasiparticles in systems
with a dominating SO interaction. Instead of naively extending the
length of the effective spin, our large-$J$ generalization preserves
the delicate entanglement between spins and orbitals. After
obtaining the classical ground states in the $J\to \infty$ limit, a
systematic $1/J$ expansion can be attained with the aid of
Holstein-Primakoff transformation. Using the large-$J$ approach, we
examine two canonical examples: the iridium and vanadium compounds
with effective spins $J=1/2$ and $J=2$, respectively. In the case of
Ir$^{4+}$ ion, the large-$J$ approach is equivalent to conventional
large-$S$ method; the classical Hamiltonian has the same form as its
quantum counterpart.

The situation for V$^{3+}$ ion is more complicated. By minimizing
the $J\to \infty$ limit of the effective Hamiltonian, we obtain a
classical phase diagram of vanadium spinels. Our result provides a
coherent explanation for the collinear and orthogonal
antiferromagnetic orders observed in spinels ZnV$_2$O$_4$ and
MnV$_2$O$_4$, respectively. The SE interaction in vanadium spinels
has rather distinct characteristics in spin and orbital channels:
while the spin exchange preserves a SU(2) symmetry, the orbital part
resembles an anisotropic 3-state Potts model. We find that the
strong spatial dependence of the orbital exchange leads to a gapped
excitation spectrum with a narrow bandwidth.

{\em Spin-orbit coupling and effective spins}. We start by
describing the general large-$J$ extension and then discuss specific
applications to Ir$^{4+}$ and V$^{3+}$ ions with $d^5$ and $d^2$
configurations, respectively. In both cases, the orbital degrees of
freedom of a partially filled $t_{2g}$ triplet are described by an
effective angular momentum of length $L'=1$. The true angular
momentum is given by $\mathbf L = \alpha \mathbf L'$, where $\alpha$
of order one is the so-called covalency factor. \cite{abragam70} The
SO interaction $V_{LS} = \alpha\lambda(\mathbf L'\cdot\mathbf S)$ is
diagonalized by the eigenstates of the ``total angular momentum''
$\mathbf J = \mathbf L' + \mathbf S$. Depending on the sign of
$\alpha$, the ground state has an effective angular momentum of
length $J_{\rm eff} \equiv L'\pm S$ and is separated from the
excited states by a gap of order $\lambda$.

In order to carry out a controlled $1/J$ expansion of the effective
Hamiltonian, we first generalize $J$ to arbitrary values by
considering a fictitious ion with $N$ identical copies of $t_{2g}$
triplets. The generalized SO interaction is $V_{LS} =
\alpha\lambda\sum_{m=1}^N \mathbf L'_m \cdot\mathbf S_m$, where
$\mathbf L'_m$ and $\mathbf S_m$ are orbital momentum and spin
operators of the $m$-th $t_{2g}$ triplet, respectively. We further
assume that only states which are symmetric with respect to the $N$
degenerate $t_{2g}$ triplets are allowed physical states. In the
atomic ground state, each $t_{2g}$ triplet has an angular momentum
of length $J_m = J_{\rm eff} = L'\pm S$, where $\mathbf J_m=\mathbf
L'_m + \mathbf S_m$, and the symmetric sum $\mathbf J = \sum_{m=1}^N
\mathbf J_m$ has the maximum length $J=N J_{\rm eff}$.

Single-ion operators such as orbital projections and spins are
extended accordingly: $P_\alpha = \sum_{m=1}^N P_{m,\alpha}$ and
$\mathbf S = \sum_{m=1}^N \mathbf S_m$. To obtain their
representation in the $J=N J_{\rm eff}$ subspace, we expand its
basis $|J,J_z\rangle$ with respect to the quantum number $J_{m,z}$
of the $m$-th $t_{2g}$ triplet
\begin{eqnarray}
    |J\!, J_z \rangle = \sum_{J_{m,z}}
    \!C_{J_z,J_{m,z}}
    |J_{\rm eff}, J_{m,z}\rangle\!\otimes\!
    |J-J_{\rm eff},  J_z-J_{m,z}\rangle,
\end{eqnarray}
where $C_{J_z,J_{m,z}}$ are the Clebsch-Gordon coefficients. By
further expressing $|J_m,J_{m,z}\rangle$ in terms of $|L'_{m,z},
S_{m,z}\rangle$, one obtains  the matrix elements of operators
acting on the $m$-th $t_{2g}$ orbitals. As the $J=N J_{\rm eff}$
subspace only contains permutationally symmetric states, the overall
operator at a site is just $N$ times that of a single $t_{2g}$
triplet, e.g. $\mathbf S = N \mathbf S_m$.

Taking into account the inter-site exchange interaction, the
effective Hamiltonian of the angular momenta $\mathbf J_i$ in the
$\lambda\to \infty$ limit is given by $H_{\rm eff} \equiv
\mathcal{P} H_{\rm SE} \mathcal{P}$, where $\mathcal{P}$ is the
projection operator of the $J=N J_{\rm eff}$ manifold, and $H_{\rm
SE}$ is the spin-orbital superexchange Hamiltonian.

{\em Ir$^{4+}$ ions: $L'=1$, $S=1/2$}. The iridium ions in compounds
such as Sr$_2$IrO$_4$ are in the low spin $5d^5$ configuration with
a positive covalency $\alpha \approx 1$; \cite{abragam70} the atomic
ground state has an effective spin $J_{\rm eff} = 1/2$. By
introducing $N$ identical $t_{2g}$ replicas as described above, we
generalize the effective spin to arbitrary large values $J=N/2$.
Restricted to the lowest-energy manifold, we find $P_{xy} = P_{yz} =
P_{zx} = 2J/3$, i.e. the electrons equally populate the three
$t_{2g}$ orbitals. The angular momentum and spin operators are given
by
\begin{eqnarray}
    \mathbf S =  -\mathbf J/3, \quad \quad
    \mathbf L' = 4 \mathbf  J/3.
\end{eqnarray}
The magnetic moment of the Ir$^{4+}$ ion is $\bm \mu =
\mu_B(2\mathbf S+\alpha \mathbf L')  \approx \frac{2\mu_B}{3}
\mathbf J$. For compounds with 180$^\circ$ Ir-O-Ir bonds, the
projected Hamiltonian
\begin{eqnarray}
    H_{\rm eff} = \mathcal{J}_1\sum_{\langle ij \rangle} \mathbf
    J_i\cdot\mathbf J_j + \mathcal{J}_2\sum_{\langle ij \rangle}
    (\mathbf J_i\cdot\hat\mathbf r_{ij})(\mathbf J_j\cdot\hat\mathbf
    r_{ij})
    \label{eq:heff1}
\end{eqnarray}
is dominated by a Heisenberg isotropic exchange plus a
pesudo-dipolar interaction. \cite{jackeli08} Here $\mathcal{J}_1
\approx 4\mathcal{J}$ and $\mathcal{J}_2 \approx 2\eta\mathcal{J}$
are effective exchange constants, $\eta = J_H/U$ is the ratio of
Hund's coupling to on-site Coulomb repulsion $U$, and
$\mathcal{J}=t^2/U$ defines the overall exchange energy scale. More
importantly, we find that the large-$J$ version of the effective
Hamiltonian, and particularly the classical limit ($J \to \infty$),
has the same form as the quantum Hamiltonian first derived in Ref.
\onlinecite{jackeli08}. The large-$J$ approach in this case thus is
equivalent to the conventional large-$S$ extension. In
Sr$_2$IrO$_4$, the `weak' ferromagnetic moment accompanying the
ground-state antiferromagnetic order is explained by treating  the
spins as classical objects and taking into account the staggered
rotations of the IrO$_6$ octahedra. \cite{jackeli08} The large-$J$
approach thus provides a theoretical basis for the classical
treatment of the effective quantum  Hamiltonian.

{\em V$^{3+}$ ions: $L'=1$, $S=1$}. Vanadium ions in cubic and
spinel vanadates have a $3d^2$ configuration and a negative
$\alpha$. Consequently, the effective angular momentum minimizing
the SO interaction has a length of $J_{\rm eff} = 2$. The large-$J$
generalization leads to the following orbital projection operator
\begin{eqnarray}
    \label{eq:Pxy}
    P_{xy} = \frac{J(J-1) + J_z^2}{2(2J-1)}.
\end{eqnarray}
For projections $P_{yz/zx}$, we replace $J_z$ by $J_x$ and $J_y$,
respectively. It could be easily checked that $P_{yz}+P_{zx}+P_{xy}
= J$, which is just the total number of electrons $2N$. Similarly,
projected to the $J=2N$ subspace, the spin and orbital momentum
operators are
\begin{equation}
    \label{eq:SL}
    \mathbf S = \mathbf J/2, \quad\quad
    \mathbf L' = \mathbf J/2.
\end{equation}
The angular momentum quanta are evenly divided between the spin and
orbital channels. The V$^{3+}$ ion has a reduced magnetic moment
$\bm \mu = \mu_B(1-|\alpha|/2) \mathbf J \approx
\frac{\mu_B}{2}\mathbf J$. Due to the spin-orbital entanglement of
the $J=2N$ states, the representation of the operator product
$P_\alpha \mathbf S$ in this subspace is different from the matrix
product of individual operators. Instead, we find
\begin{eqnarray}
    \label{eq:PxyS}
    P_{xy}\mathbf S = \frac{(J-1)\,\mathbf J}{2(2J-1)} +
    \frac{J_z\mathbf J J_z}{2(J-1)(2J-1)}.
\end{eqnarray}
The cubic symmetry of the system allows us to obtain the expressions
with $P_{yz}$ and $P_{zx}$ projections by simply replacing $J_z$ by
$J_x$ and $J_y$, respectively.

The discussion so far has been general and is applicable  to
vanadium compounds with V$^{3+}$ ions in an octahedral crystal
field. We now focus on the case of spinels where the vanadium ions
form the {\em pyrochlore} lattice, a network of corner-sharing
tetrahedra. The $90^\circ$ V-O-V bonds make direct exchange  the
primary mechanism of the SE interaction.
\cite{dimatteo05,tsunetsugu03} The resulting Potts-like orbital
interaction leads to a highly anisotropic effective Hamiltonian in
both coordinate and spin spaces.

Restricted  to the $J=2N$ manifold, the effective quantum
Hamiltonian can be obtained by substituting the operator expressions
(\ref{eq:Pxy})--(\ref{eq:PxyS}) into the SE Hamiltonian for vanadium
spinels. After introducing normalized classical spins $\hat\mathbf
J_i \equiv \mathbf J_i/J$, the effective Hamiltonian in the
classical limit $J\to \infty$ becomes
\begin{eqnarray}
    \label{eq:heff2}
    & & H_{\rm eff} = \mathcal{J}_1 J^2\sum_{\langle ij \rangle}
    \bigl(\hat J_{i,\gamma} \hat J_{j,\gamma}\bigr)^2
    \nonumber \\
    & & \quad \quad + \mathcal{J}_2 J^2 \sum_{\langle ij
    \rangle} (1+\hat J^2_{i,\gamma})(1+\hat J^2_{j, \gamma})\,
    \hat\mathbf J_i\cdot\hat\mathbf J_j \\
    & & \quad \quad -\mathcal{J}_3 J^2\sum_{\langle ij \rangle}
    \bigl[1-(\hat J_{i,\gamma} \hat J_{j,\gamma})^2\bigr]\,
    \hat\mathbf J_i\cdot\hat\mathbf J_j. \nonumber
\end{eqnarray}
Here the subscript $\gamma=\gamma(ij)$ denotes the $x$, $y$ or $z$
component of the angular momentum $\mathbf J_i$ depending on whether
the nearest-neighbor bond $\langle ij \rangle$ points along the
$\langle 011 \rangle$, $\langle 101\rangle$, or $\langle 110
\rangle$ directions, respectively. To lowest order in $\eta$, the
various exchange constants are $\mathcal{J}_1 \simeq\frac{1}{16}
(1+5\eta)\mathcal{J}$, $\mathcal{J}_2 \simeq
\frac{1}{16}(1-\eta)\mathcal{J}$, and $\mathcal{J}_3 \simeq
\frac{\eta}{8}\mathcal{J}$. \cite{dimatteo05,tsunetsugu03}

{\em Classical ground state of vanadium spinels}. There has been
much experimental effort to understand the properties of spinel
vanadates $A$V$_2$O$_4$. For divalent ion $A = $ Zn, Cd, and Mg,
magnetic moments of V$^{3+}$ ions are ordered collinearly in the
ground state. \cite{lee04,reehuis03} Theoretical models based on a
large SO coupling have been proposed to explain the observed spin
and orbital ordering. \cite{ot04,dimatteo05,maitra07} However, a
recent experimental characterization of another spinel MnV$_2$O$_4$
shows an orthogonal magnetic structure of vanadium spins.
\cite{garlea08} The mechanisms that stabilize the orthogonal
magnetic order remains unclear. Based on the large-$J$ approach,
here we propose a unifying model for the magnetic and orbital order
in vanadium spinels.

The partially filled $t_{2g}$ orbitals also couple to the
distortions of the surrounding VO$_6$ octahedron. Indeed, both types
of vanadium spinels undergo a cubic-to-tetragonal structural
transition which accompanies the orbital ordering. The lattice
distortion results in a splitting of the $t_{2g}$ triplet: $V_{\rm
JT} = \sum_{\beta} \delta_\beta P_{\beta}$, where $\beta$ runs over
the three $t_{2g}$ states. Using Eq. (\ref{eq:Pxy}), the JT coupling
can be recast into
\begin{eqnarray}
    \label{eq:JT}
    V_{\rm JT} = \delta_1\,(J_x^2+J_y^2-2 J_z^2)/\sqrt{6}
    + \delta_2\,(J_x^2 - J_y^2)/\sqrt{2},
\end{eqnarray}
where $(\delta_1, \delta_2)$ transforming as a doublet irreducible
representation under the symmetry group $O_h$ denote the tetragonal
and orthorhombic distortions of the VO$_6$ octahedron, respectively.

\begin{figure}
\centering
\includegraphics[width=0.99\columnwidth]{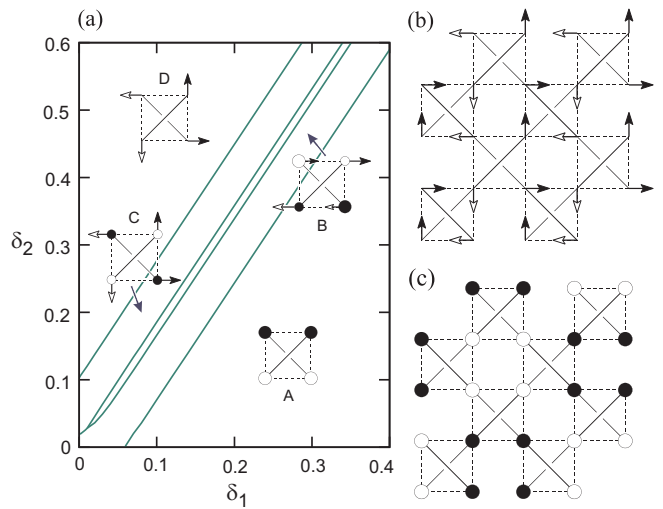}
\caption{\label{fig:phase} (a) Classical phase diagram of a
tetrahedron. Black and white circles denote $\pm J_z$ components,
respectively. $\delta_1$ and $\delta_2$ denote the uniform $E_g$ and
staggered $F_{1g}$ distortions, respectively. A narrow low-symmetric
phase with non-coplanar spins exists between phases B and C. (b) The
$\mathbf q=0$ orthogonal order and (c) the $\mathbf q=(001)$
collinear order are observed in spinels MnV$_2$O$_4$ and
ZnV$_2$O$_4$, respectively.}
\end{figure}

Structural distortions compatible with the observed tetragonal
symmetry (lattice constants $a=b > c$) in general involves $\mathbf
q=0$ phonons with symmetries $E_g$ and $F_{1g}$. \cite{chern08} The
$E_g$ distortion gives rise to a uniform anisotropy $\delta_1$,
whereas softening of rotational $F_{1g}$ phonons creates a staggered
orthorhombic distortions $\pm \delta_2$ on VO$_6$ octahedra along
the $[110]$ and $[1\bar 10]$ chains, respectively. \cite{chern08}
Classical ground state of a tetrahedron, the building block of the
pyrochlore lattice, is obtained by minimizing the sum of the
effective Hamiltonian (\ref{eq:heff2}) and JT coupling
(\ref{eq:JT}). The resulting phase diagram is shown in Fig.
\ref{fig:phase}(a).

The orthogonal magnetic order is stabilized at large values of the
staggered $F_{1g}$ distortion $\delta_2$. Interestingly, the
orthogonal structure also minimizes the $\mathcal{J}_1$ term of
$H_{\rm eff}$, which originates from the antiferro-orbital
interaction in the original SE Hamiltonian. From Eq. (\ref{eq:Pxy}),
we see that $d_{zx}$ and $d_{yz}$ orbitals are fully occupied along
the $[110]$ and $[1\bar 10]$ chains, respectively. This
antiferro-orbital ordering is consistent with the low-temperature
symmetry $I4_1/a$ of MnV$_2$O$_4$. \cite{garlea08}
We note that the inclusion of SO interaction does not seem to affect
the orbital-ordering pattern. \cite{sarkar09} This could be
attributed to the already mixed orbital states favored by a strong
local trigonal distortion in MnV$_2$O$_4$.

On the other hand, the collinear antiferromagnetic order becomes the
ground state when lattice distortion is dominated by the $E_g$ mode
with $\delta_1 > 0$, which gives rise to an easy-axis anisotropy
according to (\ref{eq:JT}). The $\mathcal{J}_2$ term in
(\ref{eq:heff2}) indicates that collinear spins parallel to the
$z$-axis have the largest antiferromagnetic coupling on
nearest-neighbor bonds lying in the $xy$ planes. The ground state
can then be viewed as a collection of antiferromagnetic Ising chains
running along the $[110]$ and $[1\bar 10]$ directions [Fig.
\ref{fig:phase}(c)], as was indeed observed in ZnV$_2$O$_4$.
\cite{lee04} The resulting orbital ordering has occupation numbers
$n_{xy} = 1$ and $n_{yz} = n_{zx} = 1/2$, also consistent with the
observed $I4_1/amd$ symmetry in ZnV$_2$O$_4$. \cite{reehuis03}
A similar staggered ordering of effective moment $\mathbf J$ is
found in cubic vanadates at large $\lambda$. \cite{horsch03}

\begin{figure}
\centering
\includegraphics[width=0.98\columnwidth]{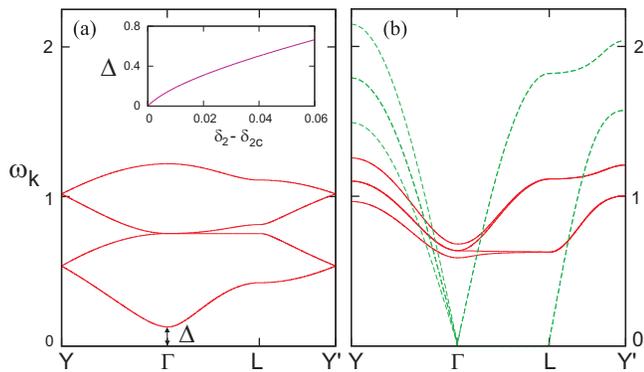}
\caption{\label{fig:spectrum} Quasiparticle spectra of vanadium
spinels with (a) $\mathbf q=0$ orthogonal  and (b)  $\mathbf
q=(001)$ collinear magnetic orders in the limit of strong spin-orbit
coupling (solid lines). Also shown for comparison is the magnon
spectrum of the collinear order without spin-orbit coupling (dashed
lines). In obtaining these spectra, we have used $\delta_1=0,
\delta_2 = 0.2\mathcal{J}$ and $\delta_1=\delta_2 = 0$ in (a) and
(b), respectively. The symmetry points in $k$-space are $\Gamma =
(0,0,0)$, $L=(\frac{1}{2},\frac{1}{2},\frac{1}{2})$, $Y=(0,1,0)$,
and $Y'=(0, 1,\frac{1}{2})$.  The energy $\omega_{k}$ is measured in
units of $\mathcal{J}\equiv t^2/U$. The inset in (a) shows the
energy gap $\Delta$ as a function of staggered distortion
$\delta_2$.}
\end{figure}

{\em Holstein-Primakoff transformation}. Once the classical ground
state is determined from the effective Hamiltonian, one can carry
out a controlled $1/J$ expansion using the Holstein-Primakoff
transformation:
\begin{eqnarray}
    J_z = J - a^{\dagger} a,\quad
    J_+ = \sqrt{2J-a^\dagger a}\,a,\quad
    J_- = J_+^\dagger.
\end{eqnarray}
The linear quasiparticle spectrum can then be obtained with the aid
of Bogoliubov transformation; the calculated spectra for the
orthogonal and collinear magnetic orders are shown by the solid
lines in Fig. \ref{fig:spectrum}.

In the case of $\mathbf q = 0$ orthogonal magnetic order, we obtain
an acoustic-like band whose gap scales as $\Delta \propto
\sqrt{\delta_2 - \delta_{2c}}$ in the vicinity of $\delta_{2c}
=0.154\mathcal{J}$. The gapless mode at $\delta_{2c}$ signals the
transition from phases D to C in Fig. \ref{fig:phase}(b). Although a
similar gapped acoustic mode ($\Delta \approx 1.5$ meV) whose origin
is attributed to the orthogonal order was indeed observed in
MnV$_2$O$_4$, \cite{garlea08} the effect of the magnetic Mn$^{2+}$
ions ($S=5/2$) remains to be clarified.  On the contrary, all
quasiparticle bands of the collinear magnetic order have an energy
gap $\Delta \approx 0.6 \mathcal{J}$ even when $\delta_1 = 0$. This
is because the staggered arrangement of frustrated bonds (parallel
spins) in the $\mathbf q=(001)$ collinear order prevents the onset
of the soft mode B shown in Fig. \ref{fig:phase}.

Also shown for comparison is magnon spectrum of the collinear order
in the absence of SO coupling (dashed lines). \cite{perkins07} The
Goldstone mode at the zone center reflects the global O(3) symmetry
of the superexchange Hamiltonian in the $\lambda = 0$ limit. The
rather narrow bandwidth of the excitation spectrum in the large
$\lambda$ limit can be attributed to the fact that quasiparticles
also carry orbital degrees of freedom whose Potts-like interactions
are static.

{\em Conclusion and outlook}. To summarize, we present a new
approach to the non-trivial problem of elementary excitations in
systems with on-site spin-orbital entanglement. For such compounds,
the ground-state magnetic order is usually accompanied by a
long-range ordering of orbital moments. The bosonic elementary
excitations resulting from deviations of the perfect order thus
carry both spin and orbital degrees of freedom. Starting with the $J
\to \infty$ limit, a systematic treatment of the elementary
excitations can be realized through the standard $1/J$ expansion.
The quasiparticle spectrum is obtained by diagonalizing the
quadratic Hamiltonian. Higher-order effects such as quasiparticle
scattering can also be studied using well developed methods from
conventional large-$S$ approach.

It is worth noting that spin-orbital entanglement could exist even
in the absence of SO coupling due to the operator structure of SE
Hamiltonian. For example, it is shown that composite spin-orbital
fluctuations in cubic vanadates lead to orbital-disordered phase and
violation of the Goodenough-Kanamori rules. \cite{oles06}
As the large-$J$ Hamiltonian $H_{\rm eff} = \mathcal{P} H_{\rm SE}
\mathcal{P}$ represents the lowest-order term in a
$\mathcal{J}/\lambda$ expansion, another important front is to
include the effects of finite $\lambda$. Higher-order terms can be
obtained following the perturbation expansion of the SE Hamiltonian.
In studying real compounds, $\lambda$ is usually of the same order
of exchange $\mathcal{J}$. Whether the excitations can be described
by the large-$J$ approach depends on the nature of the ground state.
In the presence of orbital-moment ordering, we think the large-$J$
method would be the appropriate choice.

Finally, we note that the classical effective Hamiltonian obtained
in the $J\to\infty$ limit provides a practical working model for
analyzing the experimental data. Furthermore, being a generic
approach, our method can be easily adapted to other compounds where
a large SO coupling dominates the low-energy physics.

{\em Acknowledgment}. We thank O. Tchernyshyov and Z. Hao for useful
discussions  on vanadium spinels.

\end{document}